\documentclass[aps,prl,twocolumn,showpacs,groupaddress]{revtex4}
\usepackage{amssymb}
\usepackage{graphicx}
\usepackage{dcolumn}
\usepackage{bm}
\usepackage{amsmath}

\begin{document}

\title{Nonlinear interactions with an ultrahigh flux of broadband entangled photons}
\author{Barak Dayan}
\author{Avi Pe'er}
\author{Asher A. Friesem}
\author{Yaron Silberberg}
\affiliation{Department of Physics of Complex Systems, Weizmann
Institute of Science, Rehovot 76100, Israel}

\begin{abstract}

We experimentally demonstrate sum-frequency generation (SFG) with
entangled photon-pairs, generating as many as 40,000 SFG photons
per second, visible even to the naked eye. The nonclassical nature
of the interaction is exhibited by a linear intensity-dependence
of the nonlinear process.  The key element in our scheme is the
generation of an ultrahigh flux of entangled photons while
maintaining their nonclassical properties. This is made possible
by generating the down-converted photons as broadband as possible,
orders of magnitude wider than the pump. This approach can be
applied to other nonlinear interactions, and may become useful for
various quantum-measurement tasks.
\end{abstract}


\maketitle

Entangled photons play a dominant role in quantum communication
science due to their inherent nonclassical correlations
\cite{Hong&Mandel_1985,{Jennewein@Zeilinger_PRL_2000},{Bouwmeester@Zeilinger_Nature_1997},
{Boschi@Popescu_PRL_1998},{Mattl@Zeilinger_PRL_1996}}.
Considerable efforts have been invested towards achieving
nonlinear interactions with entangled photons, since such
interactions are significant for all-optical quantum computation
and quantum metrology
\cite{Milburn_PRL_1988,{Knill@Milburn_Nature_2001},{Strekalov@Dowling_JMO_2002},{Duan@Kimble_PRL_2004}}.
Most works have focused on enhancing the nonlinear photon-photon
coupling in order to achieve conditional phase-shifts with
single-photons. Several strategies aimed at increasing the low
efficiencies of nonlinear interactions with single-photons
(typically below $10^{-10}$) were developed. One such strategy is
increasing the nonlinearity through strong photon-atom coupling in
a high-finesse cavity
\cite{Turchette@Kimble_PRL_1995,{Zubairy@Scully_PRA_2003},{Duan@Kimble_PRL_2004}};
another uses mixing of extremely weak coherent pulses with a
strong pump in a nonlinear crystal
\cite{Resch@Steinberg_PRL_2001}. Other proposed schemes rely on
enhanced nonlinearities obtained through
electromagnetically-induced transparency
\cite{Harris@Hau_PRL_1999,{Lukin@Imamoglu_Nature_2001},{Mabuchi@Doherty_Science_2002}}.
Resonantly enhanced two-photon absorption was performed with
narrowband down-converted light at power levels that approached
the entangled-photons regime, demonstrating a nonclassical
departure from quadratic intensity-dependence
\cite{Georgiades@Kimble_PRL_1995}. At lower power-levels, where
down-converted light can be considered as composed of separated
photon-pairs, a completely linear intensity-dependence of
two-photon absorption and SFG is predicted
\cite{{Banacloche_PRL_1989},{Javanainen@Gould_PRA_1990},{Ficek@Drummond_PRA_1991},{Georgiades@Kimble_PRA_1999}}.
Such a linear dependence can be shown to break the Cauchy-Schwartz
inequality for classical fields \cite{McNeil@Gardiner_PRA_1983}.\

Here we propose an alternative approach towards achieving
nonlinear interactions with entangled photons, which does not
focus on increasing the nonlinear coupling, but rather on
obtaining an ultrahigh flux of entangled photon-pairs. In our
setup we generate a flux of about $10^{12}$ entangled-pairs per
second, a flux that corresponds to a classically-high power level
of $0.3 \: \mu W$. This flux is orders of magnitude greater than
is typically utilized in quantum-optics experiments, which are
usually limited to electronic detection rates.\

The key element in achieving such a high photon-flux while
maintaining its nonclassical properties, is generating the
entangled-pairs as broadband as possible. The physical reason for
this is that the arrival times of the two photons are correlated
to within a timescale $\tau$ inversely proportional to their
bandwidth $\Delta_{DC}$ \cite{Hong&Mandel_1985}. Thus, the maximal
flux $\Phi_{max}$ of down-converted photons that can still be
considered as composed of distinct photon-pairs scales linearly
with the bandwidth
\cite{Javanainen@Gould_PRA_1990,{Dayan@Silberberg_PRA_2004}}:

\begin{eqnarray}
\label{LinQSFG_eq1} \Phi_{max}\approx\Delta_\textsc{dc} \ .
\end{eqnarray}

This maximal flux, which corresponds to a mean spectral
photon-density of $n=1$ (one photon per spectral mode), is the
crossover point between the classical high power regime and the
nonclassical low power regime.\

A complete quantum-mechanical analysis
\cite{Dayan@Silberberg_PRA_2004} shows that the rate of SFG events
with correlated pairs (i.e. photons generated from the same pump
mode) indeed includes a term that depends linearly on the
intensity:

\begin{eqnarray}
\label{LinQSFG_eq2} R_c\propto \Delta_\textsc{dc}\left(
n^2+n\right) \ .
\end{eqnarray}

The dependence on $\Delta_\textsc{dc}$ indicates that the
probability for up-conversion is inversely proportional to the
temporal-separation between the photons. This dependence on the
bandwidth is not unique to entangled photons; the rate of SFG
induced by classical pulses demonstrates the same dependence on
their bandwidth, for the same reason. It is only the linear
dependence on $n$ in Eq. \ref{LinQSFG_eq2} that is unique to
entangled photons. Note that all up-converted correlated pairs
produce SFG photons back at the wavelength of the pump.
Accordingly, in Eq. 2 it was assumed that the pump bandwidth
$\delta_p$ is completely included within the bandwidth that is
phase-matched for up-conversion in the nonlinear crystal
$\delta_\textsc{uc}$.\

Apart from the SFG events that are generated by correlated pairs
there is always a classical background noise due to "accidental"
SFG of uncorrelated photons. Unlike correlated pairs, uncorrelated
pairs can be up-converted to any wavelength in the range of
$2\Delta_\textsc{dc}$ around the pump frequency. The actual rate
of such SFG events is therefore proportional to the available
up-converted bandwidth $\delta_\textsc{uc}$
\cite{Dayan@Silberberg_PRA_2004}:
\begin{eqnarray}
\label{LinQSFG_eq3} R_{u.c.}\propto \delta_\textsc{uc}\: n^2 \ ,
\end{eqnarray}
Like in many other detection-schemes, this expression simply
measures the background noise that falls within the spectrum of
the 'receiver'.\

To clarify the role of entanglement in the SFG process, let us
first consider the narrowband case, where only one pair of signal
and idler modes are involved in the down-conversion process.
Assuming a low down-conversion efficiency that yields a mean
spectral photon density of $n \ll 1$, the state $|\: \psi \:
\rangle$ of the down-converted light can be described by:
\begin{eqnarray}
\label{01} |\: \psi \: \rangle \approx M \: | 0 \rangle +
\sqrt{n}\:| 1 \rangle_s | 1 \rangle_i \ ,
\end{eqnarray}
with $M\sim1$. The subscripts $s,i$ denote the signal and idler
modes, whose frequencies sum to the pump frequency. Note that the
spectral width of the modes can be defined as the spectral width
of the pump mode $\delta_p$ (this means we quantize the fields
according to the longest relevant time scale, which is the
coherence-time of the pump, and so the number of photons is
defined over a time-period of $\delta_p^{-1}$
\cite{Huttner@Ben-Aryeh_PRA_1990}). The state $|\: \psi \:
\rangle$ clearly describes an entanglement between the signal and
idler modes, which are both either in the vacuum state $| 0
\rangle$ or in the one-photon Fock state $| 1 \rangle$. It is
quite straight-forward to show that this entanglement is the
source for the linear term in Eq. \ref{LinQSFG_eq2}. When the
down-converted bandwidth is significantly larger than the pump
bandwidth, time and energy entanglement is created between the
photons. This actually means that the photon-pairs can be
generated in $N=\Delta_\textsc{dc}/\delta_p$ different entangled
mode-pairs:
\begin{eqnarray}
\label{many01} |\: \psi \: \rangle \approx M \: | \: 0 \rangle +
\sum_{j=1}^{N}\sqrt{n}\:| 1 \rangle_{s_j} | 1 \rangle_{i_j} \ .
\end{eqnarray}
Although the phase of each signal or idler mode separately is
inherently uncertain, the excitations of these mode-pairs are
mutually coherent, all having a combined phase that is shifted by
$\pi/2$ from the pump phase. Thus, the SFG probability-amplitudes
induced by all these pairs add coherently, resulting in an
amplification by $N^2$ of the correlated SFG rate, compared to the
narrowband case. Intuitively speaking, one factor of $N$ comes
from having $N$ more photon-pairs, and the second one comes from
the fact that the temporal separation between the photons of each
pair is smaller by a factor of $N$. On the other hand, The SFG
induced by up-conversion of uncorrelated pairs is summed
incoherently, resulting in an enhancement only by a factor of $N$.
Indeed, by dividing Eqs. \ref{LinQSFG_eq2} and \ref{LinQSFG_eq3},
we obtain that the ratio between the correlated and uncorrelated
rates is:
\begin{eqnarray}
\label{LinQSFG_eq4} \frac{R_c}{R_{u.c.}}\approx \frac{
\Delta_\textsc{dc}}{\delta_\textsc{uc}}\left(\frac{n+1}{n}\right)
\leq N \: \left(\frac{n+1}{n}\right) \ ,
\end{eqnarray}
\noindent where the inequality results from the condition of
$\delta_\textsc{uc}\geq\delta_p$. As is evident, this ratio is
decreased when the up-converted bandwidth is larger than the pump
bandwidth, hence the importance of performing a narrowband SFG.
Note that this gain of the correlated process over the
uncorrelated one holds at high-powers as well, and affects not
only SFG, but any other nonlinear mixing between the signal and
idler fields (e.g. two-photon absorption
\cite{{Georgiades@Kimble_OL_1996},{Dayan@Silberberg_PRL_2004}}).\

It is important to point out that while the gain described in Eq.
\ref{LinQSFG_eq4} results from correlations, such correlations do
not necessarily require entanglement. Coherent correlations
between two broadband fields are the basic principle of all
spread-spectrum communication schemes
\cite{Viterbi_1995,{Peer@Friesem_JLT_2004}}. In fact, without the
nonclassical linear term, Eqs. \ref{LinQSFG_eq2}-\ref{LinQSFG_eq3}
and \ref{LinQSFG_eq4} are identical to the equations for the
signal and noise in spread-spectrum communication performed with
shaped classical pulses \cite{Zheng@Weiner_OL_2000}; one has only
to replace $\Delta_\textsc{dc}$ with the bandwidth of the pulses
and $\delta_p$ with the spectral resolution of the phase
modulations performed on the pulses. However, while coherent
pulses can be easily shaped to exhibit spectral phase and
amplitude correlations, such correlations between Fock states are
inherently nonclassical and imply entanglement. This entanglement
is manifested in the fact that the gain in Eq. \ref{LinQSFG_eq4}
is stronger by a factor of $(n+1)/n$ than is classically
achievable, due to the additional, nonclassical linear term in the
correlated SFG process. At high power levels the linear term
becomes negligible and the correlations between the fields are
almost identical to the correlations that can be obtained by
shaping classical pulses. This is not to say that the increased
precision of the correlations (i.e. the squeezing) vanishes at
high powers, yet its effect on the SFG process becomes negligible.
All previous experiments that involved SFG with down-converted
light
\cite{Abram@Dolique_PRL_1986,{Kim@Shih_PRL_2001},{Dayan@Silberberg_QPH_2003}}
were performed at power levels that greatly exceeded $\Phi_{max}$,
where the intensity dependence was completely quadratic. The
correlation effects observed in these experiments are well
described within the classical framework, and are identical to
those demonstrated with shaped classical pulses
\cite{Zheng@Weiner_OL_2000,{Meshulach@Silberberg_Nature_1998}}.\

To conclude this discussion we summarize that all the nonclassical
properties exhibited in SFG of down-converted light result from
the part of the process that has a linear intensity dependence.
This part exists due to the entanglement between signal and idler
modes, and becomes negligible in high powers. Although time and
energy entanglement is not required for the nonclassical behavior,
it amplifies its effect with respect to the classical,
uncorrelated SFG process.\

Despite the fact that the SFG process exhibits a linear intensity
dependence at $n \ll 1$, it is nonetheless a two-photon process.
Thus, a random loss of either a signal or an idler photon is
equivalent to the loss of the entire pair, and will lead to a
'classical' quadratic reduction of the SFG signal. We expect
therefore, two different intensity-dependencies of the SFG
process: linear dependence, when the two-photon production rate is
changed; and quadratic dependence, when the entangled photons flux
is attenuated by any linear-optics mechanism.\

\begin{figure}[tb]
\begin{center}
\includegraphics[width=8.6cm] {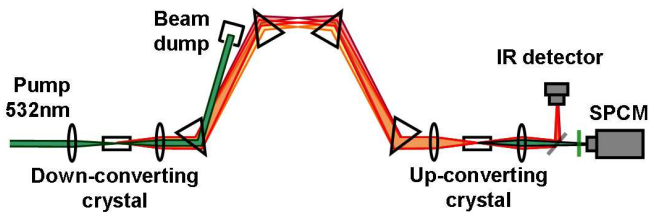}
\caption{\label{LinQSFG1} Experimental layout. Entangled photons
generated by down-conversion of a pump-laser in one crystal are
imaged through a set of four dispersion-prisms onto a second
crystal to generate the SFG photons. The entangled-photons beam is
separated form the SFG photons by an harmonic separator mirror and
its power is measured by an InGaAs detector. The SFG photons are
further filtered by $532\:nm$ line-filters and are counted with a
single-photon counting module.}
\end{center}
\end{figure}

In our experiment we used a single-frequency ($\delta_p\approx 5\:
MHz$ around $532\:nm$) doubled Nd:YAG laser to pump a $12\:mm$
long periodically-poled KTiOPO4 (PPKTP) crystal, generating
infra-red (IR) entangled-photons with a broad bandwidth of
$\Delta_\textsc{dc}=31\:nm$ around $1064\:nm$. According to Eq.
(\ref{LinQSFG_eq1}), this bandwidth implies a crossover flux of
$\Phi_{max}=8.2\cdot10^{12}$ photons per second, i.e. about
$1.5\:\mu W$. A similar PPKTP crystal was used for the SFG
process. The phase-matching conditions for the up-conversion
process in this crystal were tuned to obtain an up-converted
bandwidth of $\delta_\textsc{uc}\approx 100\:GHz$ around
$532\:nm$. According to Eq. (\ref{LinQSFG_eq4}), these bandwidths
ensure that the correlated SFG process dominates at any power. \

The experimental setup is depicted in Fig. \ref{LinQSFG1}.
Basically, entangled photons down-converted in one crystal were
up-converted in the other crystal to produce the SFG photons. The
entire layout was designed to maximize the interaction and
collection efficiencies. The non-critical phase-matching of the
PPKTP crystals eliminates the 'walk-off' between the
down-converted photons and the pump, and allows optimal focusing
\cite{Boyd@Kleinman_JAP_1968} of the pump on the first crystal and
of the entangled photons on the second one. The optimal focusing
and the high nonlinear coefficient of the crystals ($\sim
9\:\:pm/V$), yielded high nonlinear interaction efficiencies of up
to $10^{-7}$. Both crystals were temperature-stabilized to control
their phase-matching properties.\

For the SFG photons to be detected distinctly, any residue of the
pump had to be filtered-out from the down-converted photons by a
factor of at least $10^{-18}$. We chose to do so with a set of
four dispersion-prisms, designed for refraction at the appropriate
Brewster angle. This arrangement had two major advantages over
schemes which rely on harmonic-filters for filtering-out the pump.
First, Brewster-angle prisms enabled a very low loss of
down-converted photons. Second, this layout enabled a tunable
compensation of dispersion (mainly from the crystals), thereby
avoiding a significant reduction of the bandwidth effective for
the SFG process. The prisms were made of highly dispersive SF6
glass, in order to minimize the dimensions of our setup. The
entangled-photons beam was filtered-out from the SFG photons by an
harmonic-separator mirror, and its power was measured by a
sensitive InGaAs detector. The SFG photons were further filtered
by line-filters for $532\:nm$ and counted with a single-photon
counting module (SPCM-AQR-15 of $EG\&G$).\

In order to verify that any photon detected by the SPCM was the
result of the SFG process, and not a residue of the pump, we
destroyed the phase-matching in the second crystal by changing its
temperature, and observed the SPCM's count drop to its dark-count
($\sim 50\: s^{-1}$), even with the pump running at full power
($5\:W$). This dark-count level was subtracted from all the
subsequent measurements, which were performed with integration
times of $2-5\: s$, and with pump powers that did not exceed
$2.5\:W$.\

\begin{figure}[t]
\begin{center}
\includegraphics[width=8.6cm] {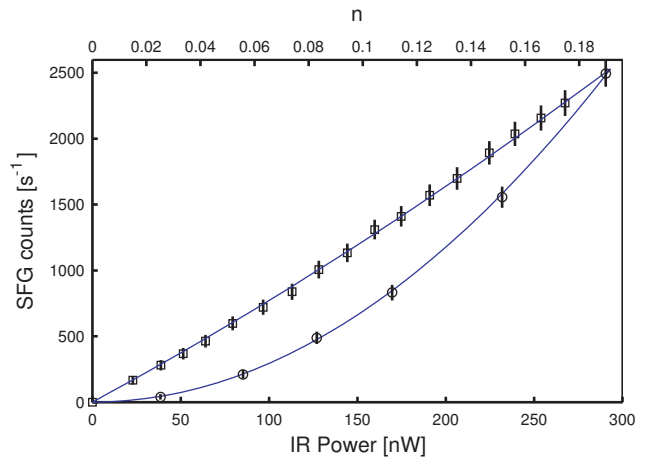}
\caption{\label{LinQSFG2} Power-dependence of SFG with entangled
photons. When the down-converted IR power is reduced by optical
attenuators, the SFG rate drops quadratically (circles:
experimental, line: quadratic fit), exactly like classical
light-sources. However, when the IR power is reduces by decreasing
the pump-laser power, thus reducing the photon-pairs production
rate, the SFG rate decreases in a close-to-linear manner, as
quantum-mechanically predicted (squares: experimental, line:
calculated $\alpha(n+n^2)$ , with $\alpha$ chosen to fit). The
log-log slope of these measurements varies from $1.01$ at the
lowest powers to $1.14$ at the highest powers. }
\end{center}
\end{figure}

We measured the power-dependence of the SFG process in two ways:
one, by attenuating the entangled-photons beam with optical
attenuators, and the other, by reducing the power of the pump
laser. Fig. \ref{LinQSFG2} depicts the results of these two
measurements. As expected, attenuation resulted in a classical,
purely quadratic decrease of the SFG counts. However, reducing the
entangled-pairs flux by decreasing the pump power resulted in a
nearly linear decrease of the SFG signal. The slight deviation
from linearity at high powers is in complete agreement with our
calculations, validating that the entire measurement was performed
in the regime of $0<n<0.185$, i.e. well below the crossover flux.
The maximal SFG count in this experiment was about $2500 \
s^{-1}$. Taking into account that the collection efficiency of the
SFG photons (limited by the transmission of the filters and the
detector's efficiency) was about $6\%$, the number of generated
SFG photons actually reached $40,000\:s^{-1}$, a flux that was
visible even to the naked eye.\

While our approach still suffers from a low nonlinear efficiency,
and thus is not readily applicable for quantum computation, it
nonetheless introduces new capabilities for quantum metrology,
since it reveals the nonclassical correlations between entangled
photons with fidelity that exceeds that of electronic
coincidences. Specifically, the SFG process simultaneously
measures both the time-difference and energy-sum of the photons,
without measuring their individual energies or times of arrival.
The large gain of $N=\Delta_\textsc{dc}/\delta_\textsc{uc}$ over
the uncorrelated SFG rate depends directly on this effect, which
implies that narrowband SFG of time and energy entangled photons
automatically rejects coincidences of non-entangled photons. This
valuable property is unattainable with electronic coincidence
detection, since electronic photo-detectors detect the actual
arrival time of each photon separately, and therefore must be as
broadband as the photons themselves.\

Our work points at the possibility to maintain the nonclassical
properties of entangled photons even at classically-high powers by
utilizing broadband, continuous down-conversion
\cite{LinQSFG_pulsed_note}. We believe this ability to perform
nonlinear interactions with ultrahigh fluxes of broadband
entangled photons holds promise in quantum-measurement science, in
particular for phase-measurements at the Heisenberg limit
\cite{Holland@Burnett_PRL_1993}.\


\begin{thebibliography}{35}
\bibitem[{Lin()}]{LinQSFG_pulsed_note}
\bibinfo{note}{With pulsed down-conversion, the rate of the photon pairs is
  practically limited by the repetition rate of the pump pulses.}
\end{thebibliography}
\end{document}